\documentclass{mn2e}
\usepackage{natbib}
\usepackage{graphicx}
\usepackage{pslatex}

\begin{document}

\title{Hot and cold bubbles in M87}

\author[C.R. Kaiser]{Christian R. Kaiser\thanks{email:
crk@astro.soton.ac.uk}\\ Department of Physics \& Astronomy,University
of Southampton, Southampton SO17 1BJ}

\maketitle

\begin{abstract}
The X-ray data obtained with {\it XMM-Newton\/} is used to investigate
the complex structure of the gas in the atmosphere of the Virgo
cluster around M87.  We construct a simple model for the temperature
and density distribution. This model implies that the cumulative mass
of the cluster gas is a power-law of its entropy index, $kT n^{-2/3}$,
similar to the Hydra cluster. This supports the idea that such
power-laws are a direct consequence of gas cooling in a gravitational
potential. In the cluster atmosphere hot bubbles of gas injected by
the AGN are rising buoyantly. We estimate the age of these structures
from the synchrotron radio data and find that this `radiative age' is
consistent with the estimated dynamical timescale. However, this
requires a spatial separation of the relativistic particles from the
magnetic field. The age estimates suggest an activity cycle of the AGN
in M87 of roughly $10^8$\, years. We show that the largest radio
structures are consistent with  being the remnants of buoyant bubbles
injected by the AGN during an even earlier activity cycle. The wakes
behind the currently rising hot bubbles uplift cold gas from the
cluster centre. Using a simple model for the trajectory of the cold
gas, we demonstrate that the observations by {\it XMM-Newton\/} of a
mix of hot and cold gas in the cluster atmosphere in the vicinity of
the radio structure can be explained in this scenario. This may also
explain the ridges of enhanced X-ray emission from cold gas observed
with {\it CHANDRA\/}.
\end{abstract}

\begin{keywords} 
cooling flows -- clusters: individual: Virgo -- galaxies: active --
galaxies: individual: M87
\end{keywords}

\section{Introduction}

The X-ray data from the {\it CHANDRA\/} and {\it XMM-Newton\/}
satellites have shown that the hot gas in galaxy clusters is not the
multiphase gas envisaged in the standard cooling flow picture
\citep[for a review see][]{bmc02}. The gas is clearly radiatively
cooling but this does not lead to distributed mass dropout and the
cooling must be interrupted at least occasionally by heating from
non-gravitational energy sources
\citep[e.g.][]{vb01,vsf02,kb02}. Recently, heating by the outflows
from AGN in the cluster centres has attracted much attention
\citep[e.g.][]{rhb00,rb02,bk01nat,ba02}. This process is complicated
and the exact details are still subject to debate. It is therefore of
crucial importance to investigate clusters containing AGN in as much
detail as possible. Because of its proximity, M87 in the Virgo cluster
is the ideal target for such studies.

X-ray data from {\it ROSAT\/} \citep{bnbf95} in combination with low
frequency radio observations \citep[][hereafter OEK00]{oek00} have
already shown that the outflow from the AGN in M87 significantly
influences the gas in the Virgo cluster \citep{bnbf95,hob99}. In this
system it has been suggested that hot, rarefied plasma bubbles
injected by the AGN are rising buoyantly in the cluster atmosphere,
uplifting in their wake colder material from the cluster centre
\citep[][hereafter CBK01]{cbkbf00}. More recent observations with {\it
XMM-Newton\/} (\citealp{bbk01,bsb01,mp01,mg01,gm02};
\citealp[][hereafter MBF02]{mbf02}; \citealp[][hereafter M02]{sm02})
and with {\it CHANDRA\/} \citep[][hereafter YWM02]{ywm02} confirm the
presence of a mixture of hot and cold gas in the regions influenced by
the AGN outflow (MBF02, M02, YMW02). Thus not radiative cooling but
the activity of the AGN creates a multi-phase medium in the cluster
atmosphere: (i) A relatively hot component representing the bulk of
the gas which would be present even without the AGN, (ii) even hotter
bubbles injected by the AGN and buoyantly rising in the gravitational
potential of the cluster and (iii) cold bubbles uplifted from the
cluster centre in the wake of the hot bubbles.

In this paper we show that the X-ray and radio observations of M87 and
the surrounding Virgo cluster are consistent with an atmosphere
containing all three of the above components. We start by relating the
idea of hot and cold bubbles to the radio and X-ray observations in
Section \ref{morph}. In Section \ref{fitatm} we will use the X-ray
data to construct a model for the temperature and density distribution
of the cluster atmosphere, component (i). We also show that the
cumulative mass of this gas is a power-law of its entropy index as
suggested for cooling cluster atmospheres by \citet{kb02}. Section
\ref{hot} derives an upper limit for the age of the buoyant hot
bubbles, component (ii), from their radio synchrotron emission and
shows that this is consistent with their estimated dynamical age. We
show that this requires the spatial separation of the relativistic
electrons from the magnetic field. In Section \ref{cold} we develop a
simple model to determine the trajectory of cold gas clouds, component
(iii), uplifted by the rise of the hot bubbles. The results from this
model are consistent with the {\it XMM-Newton\/} data. Finally, we
summarise our conclusions in Section \ref{conc}. Throughout we assume
the distance of the Virgo cluster to be 15.9\,Mpc \citep{jt91}.

\section{Morphology}
\label{morph}

Radio maps of the central regions of the Virgo cluster show three
distinct features. Following YWM02 we refer to these as the bright
`inner' radio lobes extending to about 30\,arcsec ($\sim 2.5$\,kpc)
from the cluster centre, the `intermediate' radio structures at
roughly 3\,arcmin ($\sim 15$\,kpc) and the `outer' radio halos
reaching out to 8\,arcmin ($\sim 40$\,kpc). The inner radio lobes are
the well-known, bright structures in the immediate surroundings of the
radio jet and the radio core. The intermediate radio structure to the
east of the cluster centre shows a distinct torus or `ear'-shape
structure connected to the inner lobes by a bridge of radio
emission. The entire structure resembles the `mushroom' cloud caused
by the buoyant rise of a bubble of hot gas in the gravitational
potential of the cluster (CBK01). To the west and south-west of the
centre the morphology is less clear, but OEK00 argue that the
observations indicate a torus similar to that on the eastern side
which is somewhat disrupted. In the following study of the hot buoyant
bubbles and colder bubbles uplifted in their wake, we concentrate
mainly on this intermediate structure. Finally, the outer halos give
the appearance of two almost circular regions with diameters of
4\,arcmin ($\sim 40$\,kpc) superimposed on the inner radio structures.

The high resolution X-ray {\it CHANDRA\/} observations of YWM02 reveal
ridges of enhanced emission following the intermediate radio
structures remarkably well. The X-ray ridge on the eastern side
extends to at least the centre of the radio torus, roughly 3.3\,arcmin
from the cluster centre. In the south-west, the X-ray ridge appears to
extend somewhat further. YWM02 also show that the temperature of the
gas in the region of the X-ray ridges is significantly lower than in
their surroundings. They demonstrate that the X-ray spectrum of the
ridges extending to 3.3\,arcmin is better fitted by a model assuming a
mixture of hot and cold gas compared to a single temperature
model. Based on the lower resolution {\it XMM-NEWTON\/} observations,
M02 argues that two temperature models provide a better fit to the
data out to about 6\,arcmin from the cluster centre. 

In this paper we interpret the torus structures observed in the radio
on intermediate scales as buoyant bubbles containing hot gas. Bubbles
of cold gas from the cluster centre are uplifted in the wake of the
hot bubbles. These cold bubbles could be responsible for the observed
ridges of enhanced X-ray emission and will lead to a mixture of hot
and cold gas in the vicinity of the intermediate radio structure.

\section{Fitting the cluster atmosphere}
\label{fitatm}

\begin{figure}
\centerline{
\includegraphics[width=8.45cm]{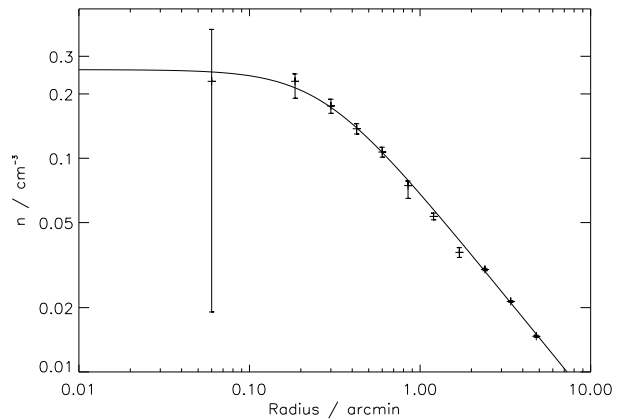}}
\caption{Model fit to the density distribution of the atmosphere of
the Virgo cluster. Data points are taken from MBF02. The solid line
shows the $\beta$-model used here.}
\label{densfit}
\end{figure}

For our investigation of the properties and evolution of the hot and
cold bubbles of gas in the surroundings of M87 in the Virgo cluster,
we require a model for the cluster atmosphere. The best data for this
purpose is provided by the X-ray observations of the Virgo cluster
obtained with {\it XMM-Newton\/}. Temperature profiles of the cluster
atmosphere can be obtained by fitting single or double temperature
MEKAL models to the data (MBF02, M02). Using spectral deprojection
techniques, the density distribution of the gas can also be determined
(MBF02). Both methods require the averaging of the observational data
over concentric annuli around the cluster centre. Thus the temperature
and density are given only at a limited number of radii (see data
points in Figure \ref{densfit}). 

In this paper we are mainly interested in the properties of the
cluster atmosphere at radii $< 6$\,arcmin. Out to this radius the {\it
XMM-NEWTON\/} observations suggest the presence of a mixture of hot
and cold gas in the cluster atmosphere (M02). We assume that the
cluster gas within this radius has been affected by the AGN
activity. Within this region we fit the density distribution with a
$\beta$-model,
\begin{equation}
n=\frac{n_0}{\left[ 1 + \left( r/a_0 \right)^2 \right] ^{1.5
\beta}},
\label{density}
\end{equation}
where $r$ is the radius from the cluster centre. The free parameters
$\beta$, $a_0$ and $n_0$ are determined by comparison of the model
with the data. We use the density points given by MBF02 which we
convert from electron density to gas density using $n = 21/11 n_{\rm
e}$.

Figure \ref{densfit} shows a comparison of the data with the
$\beta$-model for the density with $n_0=0.26$\,cm$^{-3}$, $a_0
=0.27$\,arcmin and $\beta =0.33$. We did not attempt a formal fit of
the model to the data as this would require calculating the X-ray
emission predicted by the model, folding this with the telescope
response and then analysing the model emission in the same way as the
observations. The $\beta$-model shown here is in reasonable agreement
with the data and our conclusions are not affected by the exact choice
of the model parameters. MBF02 also fit the density points with two
rather than a single $\beta$-model. We prefer the simpler single
$\beta$-model here as it provides a reasonable description of the data
within 6\,arcmin of the cluster centre.

\begin{figure}
\centerline{
\includegraphics[width=8.45cm]{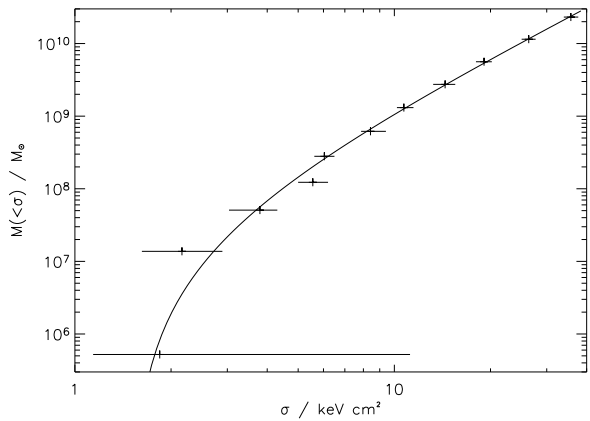}}
\caption{Plot of the cumulative mass of gas with entropy index below
$\sigma$ as function of $\sigma$. Data points are derived from the
density and temperature data given by MBF02. The error bars for
$\sigma$ result from the uncertainty in the density and temperature
measurements. The solid line shows the best fit power-law of the form
given in equation (\ref{power}).}
\label{entropyfit}
\end{figure}

The entropy index of the cluster gas is given by $\sigma = kT
n^{-2/3}$. Using the density and temperature values of MBF02 we
calculate $\sigma$ at 11 radii inside 6\,arcmin from the cluster
centre. Integrating the $\beta$-model for the density distribution, we
can also determine the mass of the gas inside these radii. Figure
\ref{entropyfit} shows the resulting plot of the cumulative gas mass,
$M(<\sigma)$ as function of $\sigma$. In the Hydra cluster
$M(<\sigma)$ is observed to follow a simple power-law relation with
$\sigma$ of the form \citep{kb02}
\begin{equation}
M(<\sigma) = A \left( \sigma - \sigma _0 \right)^{\epsilon},
\label{power}
\end{equation}
with 
\begin{equation}
A=\frac{M(<\sigma _{\rm max})}{\left( \sigma _{\rm max} - \sigma _0
\right)^{\epsilon}}.
\end{equation}
Here $\sigma _0$ is the lowest value of the entropy index of gas at
the centre of the cluster and $\sigma _{\rm max}$ is the largest
entropy index of gas contributing to the total gas mass $M(< \sigma
_{\rm max})$. The solid line in Figure \ref{entropyfit} shows the
power-law described in equation (\ref{power}) providing the best fit
to the data when taking into account the uncertainties in the entropy
index of the gas, $\sigma$. We conclude that equation (\ref{power})
allows a good approximation of the data within 6\, arcmin of the
cluster centre for $\epsilon = 2.3$, $\sigma _0=1.5$\,keV\,cm$^2$,
$\sigma _{\rm max}=40$\,keV\,cm$^2$ and $M(<\sigma _{\rm max})=2.5
\times 10^{10}$\,M$_{\odot}$. \citet{kb02} pointed out that, once
established, radiative cooling of the cluster gas with the
cooling function presented in Figure 9-9 of \citet{bt87} preserves
the power-law form of $M(<\sigma)$ and may indeed create a power-law
distribution in the first place. The observations of the hot
atmosphere in the Virgo cluster is consistent with this view.

\begin{figure}
\centerline{
\includegraphics[width=8.45cm]{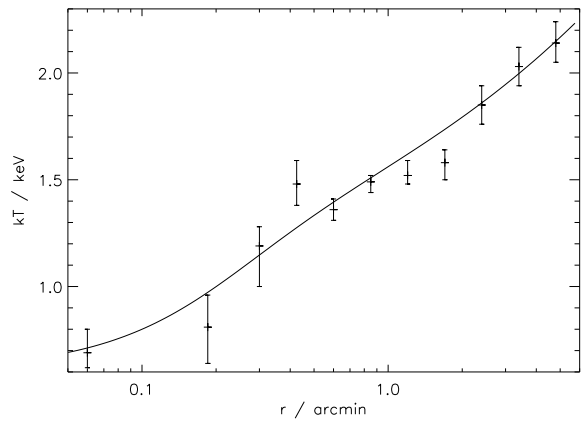}}
\caption{Comparison of the gas temperature (solid line) resulting from
the combination of the $\beta$-model of the gas density distribution
(Figure \ref{densfit}) and the power-law relation between
cumulative gas mass and entropy index (Figure \ref{entropyfit}) with
the observations of MBF02 (crosses with error bars).}
\label{temperaturefit}
\end{figure}

Combining equations (\ref{density}) and (\ref{power}) allows us to
express the entropy index as a function of radius,
\begin{equation}
\left( \sigma - \sigma _0 \right) ^{\epsilon} = \frac{4 \pi \mu m_{\rm p}n_0 a_0^3}{A} \int _0^{r/a_0} \frac{x^2}{\left( 1 + x^2 \right)^{3/2 \beta}}  \, dx,
\end{equation}
where $m_{\rm p}$ is the mass of a proton and $\mu$ is the mean
relative atomic weight of the cluster gas. The temperature of the
cluster gas is then given by $kT= \sigma n^{2/3}$. Figure
\ref{temperaturefit} compares the prediction from the models with the
measurements. Although this is not a model fit, the agreement is
reasonably good. For ideal gas conditions, we can also find an
expression for the gas pressure of the cluster gas, $p=\sigma
n^{5/3}$.

Assuming that the hot cluster atmosphere is in hydrostatic equilibrium
within the gravitational potential of the dark matter halo of the
Virgo cluster, the gravitational acceleration is given by
\begin{equation}
g = - \frac{1}{\mu m_{\rm p} n} \, \frac{{\rm d} p}{{\rm d}r}.
\label{acc}
\end{equation}
It is now straightforward to show that
\begin{equation}
g= n^{2/3} r \left\{ \frac{5 \beta \sigma}{\mu m_{\rm p} a_0^2 \left[1 + \left( \frac{r}{a_0} \right)^2 \right]} - \frac{4 \pi}{\epsilon A} \left( \sigma - \sigma _0 \right) ^{1-\epsilon} r n \right\}.
\end{equation}

\section{Hot bubbles from the AGN}
\label{hot}

\subsection{Intermediate radio structures}

We adopt the idea of CBK01 that the intermediate radio structure
resembling a torus observed to lie eastwards of the currently active
AGN in M87 (OEK00) is a bubble of rarefied plasma rising buoyantly in
the colder cluster atmosphere. The situation on the western side of
the AGN is less clear, but a similar buoyant structure may exist on
this side as well. 

The ear-like radio structure on the eastern side extends over roughly
3.2\,arcmin ($\sim 15$\,kpc) in a north-south direction. Identifying
this structure with a buoyant bubble containing gas significantly
hotter than the surrounding material implies that we may expect to
detect a depression of the X-ray surface brightness in this
region. Such `holes' have been detected in a number of clusters and
strongly support the recent ideas of heating of the cluster gas by AGN
outflows. Using the model for the gas distribution in the Virgo
cluster derived in the previous section, we can estimate the expected
depression of the X-ray surface brightness from the buoyant bubble at
this location. For a completely evacuated spherical bubble with
diameter 3.2\,arcmin at a distance of 3.3\,arcmin from the cluster
centre, the X-ray surface brightness would decrease by roughly 30\%
for a line of sight through the centre of the bubble compared to a
cluster without such a bubble. Such a strong depression is not
detected in the {\it CHANDRA\/} data. However, various effects will
make the detection of any suppression of the X-ray surface brightness
due to a buoyant bubble difficult. Firstly, because of its buoyant
motion, the hot bubble displaces and compresses cluster gas leading to
the formation of a dense shell around the bubble. Assuming the shell
thickness to be 1/10 of the bubble diameter the enhanced emission of
the dense shell roughly compensates for the emission deficit caused by
the bubble. Because of the strong dependence of the X-ray emissivity
on gas density, the bubble would show up as an X-ray {\it
enhancement\/} if the shell was thinner. Secondly, the buoyant bubble
is almost certainly not spherical because fluid instabilities will
deform it into a torus (e.g. CBK01). The observed radio structure
suggests such a toroidal geometry. The length of our line of sight
through the bubble will therefore always be shorter than the diameter
of the torus leading to a reduced effect of the bubble on the X-ray
surface brightness. Finally, there is clear evidence for cold gas in
the X-ray map at the location of the centre of the radio torus. We
will argue further on that this cold material is uplifted from the
cluster centre by the hot bubble. Its enhanced X-ray emission
complicates the detection of any X-ray depression by the hot
bubble. In all other clusters where holes in the X-ray surface
brightness are observed, the bubbles appear spherical rather than
torus-shaped. Also, in these clusters the bubbles may not be buoyantly
rising anymore which implies the absence of a dense shell around the
bubble. In fact, a close inspection of the {\it CHANDRA\/} map shows
very slight depressions of the X-ray surface brightness immediately
north and south of the torus centre (A.J. Young, private
communication). In general the deviations from perfect spherical
symmetry of the X-ray surface brightness of the gas in the Virgo
cluster makes it very difficult to detect X-ray depressions in this
highly complex region.

The detection of an X-ray hole at the location of the radio
`ear'-structure would show beyond doubt the existence of a buoyant
bubble filled with hot gas at this position. Its absence means that we
cannot prove the existence of the postulated bubble. However, the
morphology of the radio structure in combination with the work of
CBK01 strongly suggests that a hot, buoyant bubble causes the
`ear'-shaped structure to the east of the cluster centre.

CBK01 showed that a buoyant bubble will quickly reach its terminal
rise velocity, $v_{\rm t}$, defined by the balance of the buoyant and
drag forces acting on the bubble. In the case of M87 $v_{\rm t} \sim
400$\,km\,s$^{-1}$ and thus the position of the bubble as a function
of time is simply $r=v_{\rm t} t$. We cannot determine how projection
changes the apparent distance of the bubble from the cluster centre in
the radio observations of OEK00. The direction of the buoyant rise of
the hot bubbles is determined by the properties of the large-scale
distribution of gas in the cluster atmosphere rather than the
direction of the currently active jet inside the inner radio
structure. If the line connecting the eastern bubble with the cluster
centre makes an angle $\theta$ with our line of sight, then the bubble
is currently located about $15.4 \left( \sin \theta \right)^{-1}$\,kpc
from the cluster centre. If the bubble started its journey at the
cluster centre, then its age must be $t_{\rm c}=3.8 \times 10^7 \left(
\sin \theta \right)^{-1}$\,years. To demonstrate the influence of the
viewing angle $\theta$ on our results, we consider the cases $\theta
=90^{\circ}$ and $\theta = 30^{\circ}$. The latter case corresponds to
a doubling of the unprojected distance of the bubble from the cluster
centre and thus a bubble age of $t_{\rm c} = 7.6 \times 10^7$\,years.

The bubble is observed to emit synchrotron radio emission of
frequencies at least up to 10.55\,GHz \citep{rmkw96}. As CBK01 showed,
this is hard to reconcile with the usual assumption that the bubble is
filled uniformly with relativistic particles and a magnetic field
tangled on scales smaller than the bubble's size. Only if the magnetic
field is much lower than its equipartition value can the electrons
barely survive their constant energy losses for long enough to explain
the observations.

The assumption of a uniform distribution for the magnetic field in the
radio structures caused by AGN may well be too simplistic
\citep[e.g.][and references therein]{emw97}. OEK00 point out
that the entire large-scale radio structure of M87 shows filamentary
substructure that may very well indicate an inhomogeneous magnetic
field in this region. Sophisticated models of the synchrotron emission
from plasmas threaded by an inhomogeneous magnetic field have been
developed in the literature \citep[e.g.][]{pt93,pt94,emw97}. Here we
only consider the limiting case in which the radio structure is made
up of regions filled with a tangled magnetic field in pressure
equilibrium with regions devoid of any magnetic field but containing
relativistic particles. For simplicity we will also assume that the
electrons do only diffuse into the areas containing the magnetic field
but once inside do not diffuse out again. Thus, while in the
field-free regions, the electrons are subject to energy losses only
due to the adiabatic expansion of the buoyant bubble and due to
inverse Compton scattering of the Cosmic Microwave Background (CMB)
radiation. For the Lorentz factor of a relativistic electron,
$\gamma$, we can therefore write
\begin{equation}
\frac{{\rm d}\gamma}{{\rm d}t} = - \frac{\gamma}{3 V} \, \frac{{\rm
d}V}{{\rm d}t} - \frac{4 \sigma _{\rm T}}{3 m_{\rm e} c} u_{\rm c}
\gamma ^2,
\end{equation}
where $V$ is the volume of the bubble containing the relativistic
electrons, $\sigma _{\rm T}$ is the Thomson cross-section, $m_{\rm e}$
is the electron mass and $u_{\rm c}$ is the energy density of the
CMB. Because of the adiabatic expansion of the bubble during its rise,
we can replace $V$ with the pressure of the bubble which must be equal
to that of the cluster atmosphere at the location of the bubble,
\begin{equation}
\frac{{\rm d}\gamma}{{\rm d}t} = \frac{\gamma}{3 \Gamma p} \, \frac{{\rm
d}p}{{\rm d}t} - \frac{4 \sigma _{\rm T}}{3 m_{\rm e} c} u_{\rm c}
\gamma ^2,
\label{loss}
\end{equation}
where $\Gamma$ is the ratio of specific heats of the bubble
material. In the following we will assume $\Gamma = 4/3$. Using the
results of Section \ref{fitatm}, it is straightforward to numerically
solve this differential equation.

The bubble regions containing the magnetic field are in pressure
equilibrium with the cluster atmosphere. For $r=15$\,kpc, the current
position of the bubble if $\theta = 90^{\circ}$, this implies that
$B=70$\,$\mu$G. For $\theta = 30^{\circ}$, we find $r=30$\,kpc and
$B=54$\,$\mu$G. Relativistic electrons diffusing into this magnetic
field will mainly radiate at a frequency $\nu \sim \gamma ^2 e B / (2
\pi m_{\rm e} c)$, where $e$ is the elementary charge. Therefore the
eastern bubble observed at 10.55\,GHz in M87 must currently contain
electrons with a Lorentz factor $\gamma \ge 7300$ for $\theta =
90^{\circ}$ and $\gamma \ge 8400$ for $\theta = 30^{\circ}$.

\begin{figure}
\centerline{
\includegraphics[width=8.45cm]{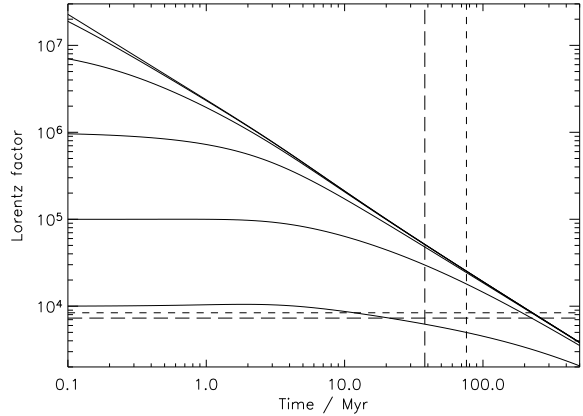}}
\caption{Evolution of the Lorentz factor for electrons inside the
buoyant bubble. Solid lines show evolutionary paths for electrons with
initial Lorentz factors equal to (from top to bottom) $10^{9}$,
$10^8$, \ldots, $10^4$. The dashed, horizontal lines show the Lorentz
factor of electrons currently required in the bubble to explain the
radio observations (Long-dashed: $\theta =90^{\circ}$, short-dashed:
$\theta =30^{\circ}$). The dashed, vertical lines show the estimated
current age of the bubble for the two viewing angles considered here.}
\label{lorentz}
\end{figure}

Figure \ref{lorentz} shows the results of solving Equation
(\ref{loss}) for the pressure profile derived in Section
\ref{fitatm}. We assume that electrons are injected into the bubble at
time $t=0$ and then passively loose energy during the rise of the
bubble. To explain the observations the electrons must clearly be
injected with Lorentz factors of several $10^4$. It is also
interesting to note that even for initially $\gamma \rightarrow
\infty$ the lifetime of the electrons is limited to about $2 \times
10^8$\,years. This is mainly caused by the inverse Compton scattering
losses. If we neglect the adiabatic term in Equation (\ref{loss}) and
set $\gamma \rightarrow \infty$ at $t=0$, we can solve directly for
$\gamma$ yielding
\begin{equation}
\gamma = \frac{3 m_{\rm e} c}{4 \sigma _{\rm T} u_{\rm c} t}, 
\end{equation}
which gives $t_{\rm max} = 3 \times 10^8$\, years for $\gamma = 8000$. 

From these considerations it becomes clear that the observed radio
emission can be explained in terms of a single injection of
relativistic electrons into the buoyant bubble. However, this almost
certainly requires a non-uniform magnetic field structure. In the case
of the eastern torus the maximum lifetime of the relativistic
electrons is consistent with the age of the structure estimated from
dynamical considerations. Even for a viewing angle to the path of the
hot bubble of $\theta =30^{\circ}$ to our line of sight, an initial
injection of relativistic electrons into the buoyant bubble is
consistent with the radio observations. However, the relativistic
electrons could easily diffuse into the cluster atmosphere if they are
not bound to the bubble by at least a weak magnetic field. The field
needed to prevent diffusion is very weak but would of course further
shorten the lifetimes of the electrons.

The alternative to the scenario sketched above is that relativistic
particles are constantly re-accelerated in the buoyant bubble. This
would require the presence of shocks and/or strong turbulence. Such
sites of ongoing particle acceleration should in principle be very
conspicuous as bright spots in the radio images. The observed
filaments could of course mark regions of particle
acceleration. However, they are not very much brighter than their
surroundings and they are more naturally interpreted as local
enhancements of the magnetic field. In any case, shocks or strong
turbulence would require a supersonic or at least chaotic fluid flow
inside the buoyant bubble. This is hard to reconcile with the subsonic
and comparatively smooth buoyant rise of the bubble in the cluster
atmosphere suggested for the formation of the torus and circular
structures.

\subsection{Outer Radio structures}

All the arguments above also apply to the relativistic electrons in
the almost circular outer radio structures enveloping the tori on
intermediate scales discussed so far. CBK01 interpreted the outer
structures as buoyant bubbles that were released from the AGN even
earlier than the torus structures and have reached their iso-entropy
surfaces. If this interpretation is correct, the outer radio
structures cannot be older than a few $10^8$\,years or otherwise we
would not detect them. This implies that the AGN in M87 must have a
duty cycle of roughly $10^8$\,years or shorter.

At the iso-entropy surface the bubbles stop rising and start
spreading. Thus the circular emission regions should be thin
(CBK01). As they must be roughly in pressure equilibrium with their
surroundings, the location of these bubbles at their iso-entropy
surface implies that their mass density must be equal to that of the
surrounding cluster medium. The density of the relativistic electrons
responsible for the observed radio synchrotron emission is orders of
magnitude lower than that of the cluster gas. Therefore the spreading
bubbles must contain a significant amount of thermal material mixed
with the relativistic particles and the magnetic field. Obviously we
cannot decide whether this thermal material was present in the bubbles
at the time their buoyant rise started or whether it was mixed in
during the rise itself. In any case, the thermal electrons embedded in
the magnetic field could lead to Faraday rotation and thus
depolarisation of the radio synchrotron emission.

\citet{rmkw96} find the polarisation of the synchrotron emission at
10.55\,GHz from the outer radio structures to exceed 70\% of the total
flux. This is close to the theoretical upper limit for the
polarisation and indicates that the synchrotron radiation suffers very
little depolarisation. We use the simple models of \citet{cj80} to
calculate the expected internal depolarisation due to the thermal
electrons inside the spreading bubbles. For a conservative estimate we
assume that the circular outer radio structures are at the same
distance ($\sim 15$\,kpc) from the cluster centre as the intermediate
tori. The $\beta$-model fit for the density of the cluster gas derived
in section \ref{fitatm} then implies an electron density of
$0.01$\,cm$^{-3}$ inside the spreading bubbles. We also assume that
the magnetic field inside the bubbles is comparable to that in the
intermediate torus structures (70\,$\mu$G). If we allow a maximum
reduction of 5\% of the fraction of polarised flux, the thickness of
the bubbles along our line of sight must be less than 1\,kpc. Although
we cannot place any other constraints on this dimension of the
bubbles, the picture of a spreading bubble is probably consistent with
a ratio of bubble thickness and bubble diameter in the plane of
spreading of 1/40. In case the spreading bubbles are further away from
the cluster centre, then both the electron density and the strength of
the magnetic field are probably lower than assumed here. This leads to
a higher upper limit on the thickness of the spreading
bubbles. Finally, it should be noted that high degrees of polarisation
do not necessarily imply the absence of thermal electrons in the
synchrotron emission region. If the structure of the magnetic field is
more complex than in the simple models used here, then large amounts
of thermal material can be present without significant depolarisation
taking place \citep{rl84}.

\section{Cold bubbles lifted by hot bubbles}
\label{cold}

Simulating the buoyant rise of a hot bubble from the AGN, CBK01 found
that cold material from the cluster centre was lifted to larger
distances from the cluster centre in the wake of the hot
bubble. Recently, MBF02 and M02 noted that the X-ray observations of
M87 obtained with {\it XMM-Newton\/} strongly suggest the presence of
a mixture of hot and cold plasma in the region influenced by the
buoyant bubble. Here we argue that the cold component of this mixture
can be identified with bubbles of cold gas uplifted from the cluster
centre in the wake of a buoyant, hot bubble. The cold bubbles may also
form the ridges of enhanced X-ray emission detected in the high
resolution maps obtained with {\it CHANDRA\/} (YWM02).

The exact details of the trajectory of a bubble of cold gas dragged
along by the rise of a hot bubble are complicated and their
investigation require numerical simulations. Here we develop a very
simple model for the motion of the cold bubbles in the cluster
atmosphere. We assume that the passage of the hot bubble accelerates
the cold bubble instantaneously to a fraction $f \le 1$ of the
terminal rise velocity of the hot bubble, $v_{\rm t}$. Thus the cold
bubble moves outward but is decelerated because of a buoyant force,
\begin{equation}
F_{\rm b} = - V \left( \rho _{\rm a} - \rho _{\rm b} \right) g,
\end{equation}
acting on it. Here, $V$ is the volume of the cold bubble, $\rho _{\rm
b}$ and $\rho _{\rm a}$ are the mass density of the cold bubble and of
the cluster atmosphere in the vicinity of the bubble respectively and
$g$ is the gravitational acceleration exerted by the dark matter halo
(see Equation \ref{acc}). The cold bubble is surrounded by cluster gas
which would normally exert a drag force on the bubble. However, as all
the material in the wake of the hot bubble is moving with roughly the
same velocity, because it is subject to the same gravitational
acceleration, we can neglect the drag force here. This situation
changes once the cold bubble comes to rest and starts sinking back
towards the cluster centre under the influence of the buoyant
force. Now its surroundings are at rest and will exert a drag force
\begin{equation}
F_{\rm d} = \frac{1}{2} C S v^2 \rho _{\rm a},
\end{equation}
where $S$ is the surface area of the bubble on which the drag force
acts and $v$ is its velocity. The drag coefficient $C$ is of order
unity and we set it to 0.75 which was found for the hot, buoyant
bubble in the simulations of CBK01. The equation of motion for the
cold bubble is therefore
\begin{equation}
\frac{{\rm d}^2 r}{{\rm d}t^2} = g \left( 1 - \frac{\rho _{\rm
a}}{\rho _{\rm b}} \right) + \frac{3}{4} \, \frac{C}{R} \, \left(
\frac{{\rm d}r}{{\rm d}t} \right) ^2 \, \frac{\rho _{\rm a}}{\rho
_{\rm b}},
\label{motion}
\end{equation}
where we have assumed that the cold bubble is a sphere with radius
$R$. During its rise and subsequent fall, the cold cloud will be in
pressure equilibrium with the cluster atmosphere. The radiative
cooling time of the cloud material is long compared to the dynamical
timescales considered here and so the evolution of the cloud will be
adiabatic. The pressure equilibrium and the adiabatic behaviour allow
us to self-consistently determine the radius, $R$, and other
properties of the cold cloud.

\begin{figure}
\centerline{
\includegraphics[width=8.45cm]{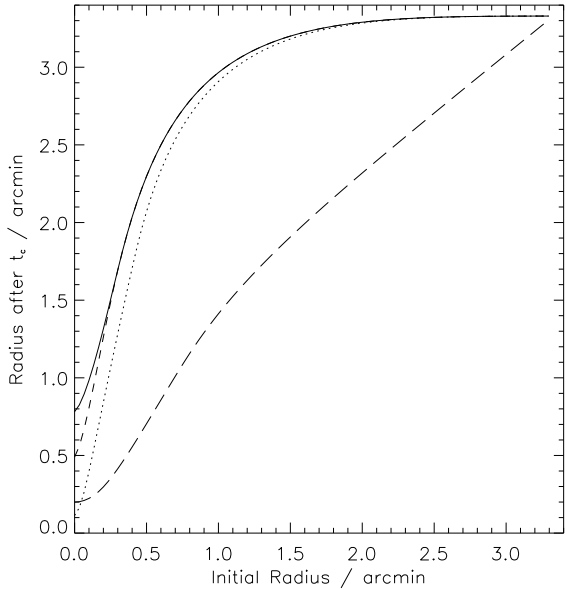}}
\caption{The position of an uplifted cold cloud at time $t_{\rm c}$
after the hot bubble started rising from the cluster centre. Solid
line: Angle of bubble motion to line of sight $\theta = 90^{\circ}$
(this implies $t_{\rm c} =3.8 \times 10^7$\,years), initial radius of
uplifted bubbles $R_{\rm s}=50$\,pc, initial rise velocity of cold
bubbles as fraction of terminal velocity of hot bubble
$f=1$. Long-dashed line: Same as solid line but $f=0.25$. Short-dashed
line: Same as solid line but $R_{\rm s}=500$\,pc. Dotted line: Same as
solid line but $\theta =30^{\circ}$, in this case $t_{\rm c}=7.6
\times 10^7$\,years.}
\label{position}
\end{figure}

Consider a bubble of gas located initially at a radius $r_{\rm s}$
from the cluster centre. For this bubble we start the integration of
Equation (\ref{motion}) at the time the hot bubble passes through
$r_{\rm s}$. At this point in time, the hot bubble accelerates the
cold bubble instantaneously to $f v_{\rm t}$. While the cold bubble is
rising, we neglect the drag force. We set the initial radius of the
cold bubbles to $R_{\rm s}$. The position of the cold bubble at time
$t_{\rm c}$ after the hot bubble started its rise in the cluster
centre, i.e. the current time, is shown in Figure \ref{position} as a
function of $r_{\rm s}$. The result depends somewhat on the angle to
our line of sight $\theta$ of the path of the cold bubbles in the wake
of the hot bubble. However, projection reduces this effect because the
longer rise time of the hot bubble compensates for the longer
unprojected distances the cold bubbles have to travel. Maybe somewhat
surprisingly the size of the uplifted, cold bubbles has a very limited
effect on their final position. However, only the contribution of the
drag force to equation (\ref{motion}) depends on the bubble size. All
cold bubbles originally located outside a radius of about 0.3\,arcmin
from the cluster centre are still rising at time $t_{\rm c}$. By
construction, no drag force acts on these clouds and so their motion
is independent of their size.

The results presented in Figure \ref{position} depend crucially on the
initial velocity of the cold clouds. If the buoyant hot cloud does not
accelerate the cold clouds to a significant fraction of its own rise
velocity, then the restructuring of the cluster atmosphere through
uplift of cold clouds is not very significant. Even for a moderate
reduction of their initial velocity ($f=0.25$), the cold clouds do not
rise very far in the cluster before starting to fall back to their
initial positions.

\begin{figure}
\centerline{
\includegraphics[width=8.45cm]{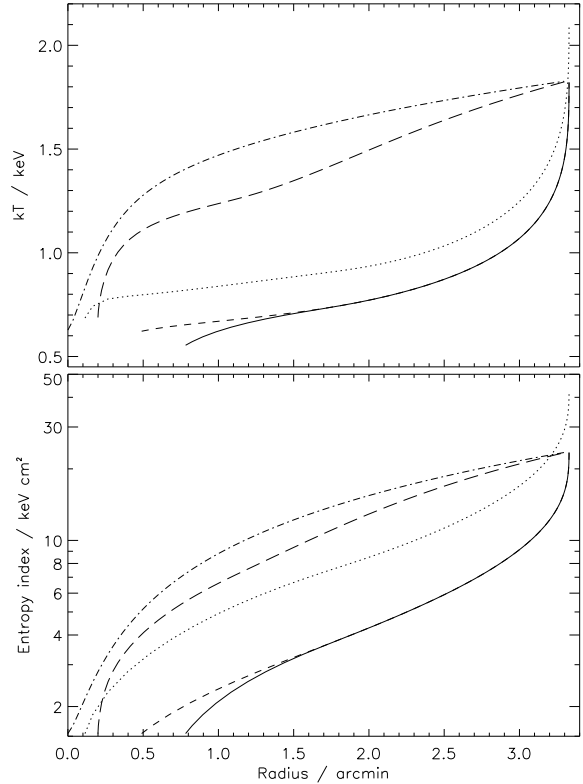}}
\caption{Temperature and entropy index distribution of the cold clouds
as a function of radius at time $t_{\rm c}$ after the hot bubble
started rising from the cluster centre. Line styles as in Figure
\ref{position}. The dot-dashed line shows the temperature and entropy
index distribution of the unperturbed cluster gas for $\theta =
90^{\circ}$.}
\label{tempent}
\end{figure}

Initially we assume that the cold clouds are part of the cluster gas
at radius $r_{\rm s}$, i.e. their temperature and density are given by
our models of the cluster atmosphere discussed in Section
\ref{fitatm}. Using the adiabatic evolution of the cold clouds we can
readily calculate the temperature and the entropy index of the cloud
material at later times. In Figure \ref{tempent} we show the
distribution of the temperature and entropy index in the unprojected
cluster atmosphere and for the uplifted cold clouds at the current
time $t_{\rm c}$. While rising, the cold clouds expand to adjust to
the lower pressure further out in the cluster. Therefore they cool
while their entropy index is of course constant. The pressure
distribution in the cluster atmosphere leads to a virtually constant
temperature of the cold clouds for most of the range in radius they
occupy within the cluster. The exact value of this temperature depends
on the projection angle $\theta$, but the flat temperature profile is
preserved independent of the values of $\theta$ or $f$. Again we find
that only for a slower initial velocity of the cold clouds this result
is significantly altered because the cold clouds do not rise very far
within the cluster atmosphere. A constant temperature of the uplifted
cold bubbles is consistent with the observational results of M02. Note
also that for strong projection, i.e. small values of $\theta$, the
temperature of the cold clouds located furthest away from the cluster
centre naturally exceeds the temperature of the unperturbed gas of the
unprojected cluster atmosphere. Deprojection would show that this
material is located at larger radii where the temperature is higher
than the projected image suggests.

The entropy index distribution of the cold bubbles is simply shifted
and stretched in the direction of increasing radius compared with the
unperturbed cluster gas. Again this is consistent with the findings of
M02. Projection has a much stronger influence on the entropy index
distribution compared to the temperature distribution. For small
values of $\theta$ we would expect a small difference between the
entropy index of the unperturbed cluster gas compared to that of the
cold bubbles. M02 find that for a given radius the entropy index of
the cold gas component in the Virgo cluster is roughly a factor 3
lower than that of the hot component. A qualitative comparison with
Figure \ref{tempent} shows that this indicates $\theta \sim
90^{\circ}$ for the cold bubbles uplifted by the hot bubble associated
with the `intermediate' radio structure in the Virgo cluster.

Note here that there is no {\it a priori\/} reason why observations of
cold clouds uplifted by hot bubbles from AGN in other clusters should
result in a similar constant temperature distribution. However, for
the specific conditions of the cluster atmosphere in the Virgo cluster
the observations of cold material associated with the radio structure
are consistent with the uplift of cold clouds described here.

\section{Conclusions}
\label{conc}

The excellent data on the hot, gaseous atmosphere of the Virgo cluster
obtained with {\it XMM-Newton\/} allows us to accurately fit the
temperature and density distribution of the cluster gas. Using these
models we find that the cumulative mass of the cluster gas,
$M(<\sigma)$, is a power-law function of the form presented in
Equation (\ref{power}) of the entropy index $\sigma = kT n^{-2/3}$. For
the Virgo cluster we find $\epsilon = 2.3$ and $\sigma _0 =
1.5$\,keV\,cm$^2$. The Virgo cluster is therefore the second cluster
after the Hydra cluster for which such a power-law has been
found. This supports the suggestion of \citet{kb02} that these
distributions arise naturally in cluster atmospheres under the
influence of radiative cooling by bremsstrahlung.

We revisit the idea that the torus structure seen to the east of the
currently active AGN in M87 is caused by the buoyant rise of a hot
bubble injected by an earlier activity cycle of the AGN. If there was
a single injection of relativistic electrons responsible for the
synchrotron radio emission into the bubble at the time it started its
buoyant rise, then the fact that it is still detectable at 10.55\,GHz
\citep{rmkw96} implies a maximum age of $3\times 10^8$\,years for this
structure. However, this requires that the relativistic electrons are
separated from the magnetic field for most of the time. The dynamical
age is of order $5\times 10^7$\,years and is therefore consistent with
the `radiative age' above. This result does not depend projection
effects.

The upper limit for the age derived here also applies to the circular
emission regions seen in the map of OEK00. If these are the remnants
of buoyant bubbles from an even earlier activity cycle as suggested by
CBK01, then the duty cycle of the AGN in M87 must be of order
$10^8$\,years. From the observations we cannot rule out a process
constantly re-accelerating relativistic particles in the radio
structures. However, their appearance and their interpretation as
slowly rising buoyant structures make such a process unlikely. We
propose that the bubbles forming the circular emission regions have
reached their iso-entropy surface in the cluster atmosphere. This
implies that they contain significant amounts of thermal material
mixed with the relativistic particles and magnetic field giving rise
to the radio synchrotron emission. We show that this scenario is
consistent with the high degree of polarisation of the observed radio
emission if the circular emission regions are thin and therefore seen
in projection.

The hot bubble drags in its wake colder material from the cluster
centre further out. We develop a simple model for the trajectory of
such cold clouds. From this model we calculate the entropy index and
temperature structure of the cold material in the Virgo cluster. The
temperature of the cold material should be roughly constant as a
function of radius over a large range of radii. The entropy index
distribution of the cluster atmosphere is shifted and stretched in the
direction of larger radii for the cold material. Both results depend
crucially on the initial velocity with which the cold bubbles start
rising in the wake of the hot bubble. If this initial velocity is
significantly lower than the rise velocity of the hot bubble, then no
flat temperature profile is found. Projection effects do not alter the
shape of the temperature or entropy index distributions. However, they
determine the absolute values of these gas properties of the cold
clouds at a given radius. Our findings do not depend significantly on
the size of the cold bubbles.

A flat temperature profile and the entropy distribution are consistent
with the temperature and entropy index distribution of the cold
component in the cluster atmosphere derived directly from the X-ray
data by M02. The flat temperature structure arises from the uplifting
of material from a range of initial radii. This material naturally has
a range of initial temperatures. The pressure distribution of the
cluster atmosphere of Virgo then leads to the observed temperature
distribution. We conclude that the mixture of hot and cold gas in the
atmosphere of Virgo can be explained by the uplift of cold gas from
the cluster centre by the buoyant hot bubbles injected by the AGN.

The Virgo cluster is probably the best studied cooling flow cluster
containing an active AGN. The superb quality of the X-ray and radio
data enable us to study the interaction of the AGN and the cluster gas
in great detail. The results presented here suggest that further X-ray
studies of other clusters containing AGN should reveal a power-law
relation between the gas mass and the entropy index of the gas and a
mixture of hot and cold gas in the regions influenced by the AGN
activity.

\section*{Acknowledgments}

The author would like to thank Dr. A.J. Young and Dr. M. Hardcastle for
useful discussions and the anonymous referee for many helpful
comments.

\bibliography{../../crk}
\bibliographystyle{mn2e}

\end{document}